\documentclass[prd,onecolumn,nofootinbib,showpacs,showkeys]{revtex4}
\newcommand\gothfamily{\usefont{U}{ygoth}{m}{n}}
\DeclareTextFontCommand{\textgoth}{\gothfamily}

\begin{document}

\title{GRAVITATION, ELECTROMAGNETISM AND COSMOLOGICAL CONSTANT\\IN PURELY AFFINE GRAVITY}

\author{{\bf Nikodem J. Pop\l awski}}

\affiliation{Department of Physics, Indiana University, Swain Hall West, 727 East Third Street, Bloomington, IN 47405, USA}
\email{nipoplaw@indiana.edu}

\noindent
{\em Foundations of Physics}\\
Vol. {\bf 39}, No. 3 (2009) 307--330\\
\copyright\,Springer Science+Business Media, LLC
\vspace{0.4in}

\begin{abstract}
The Ferraris-Kijowski purely affine Lagrangian for the electromagnetic field, that has the form of the Maxwell Lagrangian with the metric tensor replaced by the symmetrized Ricci tensor, is dynamically equivalent to the metric Einstein-Maxwell Lagrangian, except the zero-field limit, for which the metric tensor is not well-defined.
This feature indicates that, for the Ferraris-Kijowski model to be physical, there must exist a background field that depends on the Ricci tensor.
The simplest possibility, supported by recent astronomical observations, is the cosmological constant, generated in the purely affine formulation of gravity by the Eddington Lagrangian.
In this paper we combine the electromagnetic field and the cosmological constant in the purely affine formulation.
We show that the sum of the two affine (Eddington and Ferraris-Kijowski) Lagrangians is dynamically inequivalent to the sum of the analogous ($\Lambda$CDM and Einstein-Maxwell) Lagrangians in the metric-affine/metric formulation.
We also show that such a construction is valid, like the affine Einstein-Born-Infeld formulation, only for weak electromagnetic fields, on the order of the magnetic field in outer space of the Solar System.
Therefore the purely affine formulation that combines gravity, electromagnetism and cosmological constant cannot be a simple sum of affine terms corresponding separately to these fields.
A quite complicated form of the affine equivalent of the metric Einstein-Maxwell-$\Lambda$ Lagrangian suggests that Nature can be described by a simpler affine Lagrangian, leading to modifications of the Einstein-Maxwell-$\Lambda$CDM theory for electromagnetic fields that contribute to the spacetime curvature on the same order as the cosmological constant.
\end{abstract}

\maketitle

\section{Introduction}
\label{secIntro}

There exist three pictures of general relativity.
In the {\em purely affine} (Einstein-Eddington) formulation~\cite{Ein,Edd,Schr,Kij,CFK}, a Lagrangian density depends on a torsionless affine connection and the symmetric part of the Ricci tensor of the connection.
This formulation defines the metric tensor as the derivative of the Lagrangian density with respect to the Ricci tensor, obtaining an algebraic relation between these two tensors.
It derives the field equations by varying the total action with respect to the connection, which gives a differential relation between the connection and the metric tensor.
This relation yields a differential equation for the metric.
In the {\em metric-affine} (Einstein-Palatini) formulation~\cite{Pal1,Pal2,Pal3}, both the metric tensor and the torsionless connection are independent variables, and the field equations are derived by varying the action with respect to these quantities.
The corresponding Lagrangian density is linear in the symmetric part of the Ricci tensor of the connection.
In the {\em purely metric} (Einstein-Hilbert) formulation~\cite{Hilb1,Hilb2,Hilb3,Hilb4,LL2}, the metric tensor is a variable, the affine connection is the Levi-Civita connection of the metric and the field equations are derived by varying the action with respect to the metric tensor.
The corresponding Lagrangian density is linear in the symmetric part of the Ricci tensor of the metric.

Ferraris and Kijowski showed that all three formulations are dynamically equivalent~\cite{FK3a,FK3b}.
Although to each metric-affine or purely metric Lagrangian for the gravitational field and matter there corresponds the dynamically equivalent purely affine Lagrangian~\cite{Kij,FK3a,FK3b,KW}, the explicit form of such a Lagrangian is known only for few cases, including: the cosmological constant~\cite{Edd}, the Klein-Gordon~\cite{Kij}, Maxwell and Proca fields~\cite{FK1}, and barotropic fluids~\cite{KM1,KM2}.
This equivalence can be generalized to theories of gravitation with Lagrangians that depend on the full Ricci tensor and the segmental (homothetic) curvature tensor~\cite{univ,nonsym}, and to a general connection with torsion~\cite{nonsym}.

The fact that Einstein's relativistic theory of gravitation~\cite{Ein1} is based on the affine connection rather than the metric tensor was first noticed by Weyl~\cite{Weyl}.
This idea was developed by Eddington, who constructed the simplest purely affine gravitational Lagrangian, proportional to the square root of the determinant of the symmetrized Ricci tensor~\cite{Edd}.
This Lagrangian is equivalent to the metric Einstein-Hilbert Lagrangian of general relativity with the cosmological constant.
Schr\"{o}dinger elucidated Eddington's affine theory and generalized it to a nonsymmetric metric~\cite{Schr1} which was introduced earlier by Einstein and Straus~\cite{Ein2} to unify gravitation with electromagnetism~\cite{Band,Eis} (in this paper we do not attempt to unify gravitation with electromagnetism, for the review of unified field theories see~\cite{Goe}).
There also exists a quantum version of the Eddington purely affine Lagrangian~\cite{Mart1,Mart2}.

The purely affine formulation of gravity cannot use the metric definition of the energy-momentum tensor as the tensor conjugate to the metric tensor with the matter action as the generating function, since matter should enter the Lagrangian before the metric tensor is defined.
Thus matter fields must be coupled to the affine connection and the curvature tensor in a purely affine Lagrangian.
Ferraris and Kijowski found that the purely affine Lagrangian for the electromagnetic field, that has the form of the Maxwell Lagrangian with the metric tensor replaced by the symmetrized Ricci tensor, is dynamically equivalent to the Einstein-Maxwell Lagrangian in the metric formulation~\cite{FK1}.
This equivalence was demonstrated by transforming to a reference system in which the electric and magnetic vectors (at the given point in spacetime) are parallel to one another.
Such a transformation is always possible except for the case when these vectors are mutually perpendicular and equal in magnitude~\cite{LL2}.

The purely affine formulation of gravitation is not a modified theory of gravity but general relativity itself, written in terms of the affine connection as a dynamical configuration variable.
This equivalence with general relativity, which is a metric theory, implies that purely affine gravity is consistent with experimental tests of the weak equivalence principle~\cite{Wi}.
Moreover, experimental tests of the interaction between the gravitational and electromagnetic fields are restricted to the motion of quanta of the electromagnetic field, photons, in curved spacetime~\cite{Wi,Niko1}.
Therefore the predictions of purely affine gravity on how the electromagnetic field in the presence of the cosmological constant affects spacetime curvature are consistent with current observational tests of relativity.

However, the purely affine formulation of the electromagnetic field~\cite{FK1} has one defect.
In the zero-field limit, the Ferraris-Kijowski Lagrangian density\footnote{
In this paper, by the Ferraris-Kijowski Lagrangian we mean the Lagrangian density for the purely affine version of the Einstein-Maxwell theory from Ref.~\cite{FK1}, where the variables are a torsionless connection and the electromagnetic potential.
This Lagrangian density depends on the symmetric part of the Ricci tensor of the symmetric connection and on the electromagnetic field tensor.
There exists another, conceptually different, purely affine Lagrangian density of Ferraris and Kijowski for the unified field theory~\cite{FK2a,FK2b}, where an asymmetric connection with a torsion with nonvanishing trace is the only variable.
That Lagrangian depends on the symmetric part of the Ricci tensor of the connection and on the segmental curvature tensor which plays the role of the electromagnetic field tensor.
}
vanishes, making it impossible to construct the metric tensor.
The same problem arises for other forms of matter that are represented by the Ricci tensor.
For the Ferraris-Kijowski model to be physical, there must exist a background field that depends on the Ricci tensor so that the metric tensor is well-defined even in the absence of matter.
The simplest possibility, supported by recent astronomical observations~\cite{acc1,acc2,const}, is the cosmological constant.
Therefore spacetime with cosmological constant plays the role of a purely affine vacuum which should be included in constructing purely affine Lagrangians.

In this paper we study how to modify the Ferraris-Kijowski model of electromagnetism so that the metric structure in the zero-field limit of this model is well-defined.
We combine the electromagnetic field and the cosmological constant in the purely affine formulation so that the zero-field limit corresponds to spacetime with the cosmological constant alone.
In Sect.~\ref{secField} we review the field equations of purely affine gravity, generalizing the Einstein-Schr\"{o}dinger derivation~\cite{Schr} to Lagrangians that also depend explicitly on the connection.
In Sect.~\ref{secCor} we review the correspondence between the purely affine formulation of gravitation and the metric-affine and metric formulations.
Sections~\ref{secEdd} (on the Eddington Lagrangian) and~\ref{secFK} (on the Ferraris-Kijowski Lagrangian) precede Sect.~\ref{secWeak} in which we construct a purely affine version of the Einstein-Born-Infeld theory~\cite{BI,Motz,Vol} describing both the electromagnetic field and the cosmological constant.
The corresponding Lagrangian combines the symmetrized Ricci tensor and the electromagnetic field tensor before taking the square root of the determinant.
We show that this formulation is valid only for weak electromagnetic fields, on the order of the magnetic field in interstellar space, at which this Lagrangian reduces to the sum of the Eddington Lagrangian and the Lagrangian of Ferraris and Kijowski~\cite{FK1}.
In addition, we complete (in Sect.~\ref{secFK}) the proof of the equivalence of the Einstein-Maxwell and Ferraris-Kijowski Lagrangians by showing it for the case when the electric and magnetic vectors are mutually perpendicular and equal in magnitude.

In Sect.~\ref{secEddFK} we assume that the sum of the Eddington and Ferraris-Kijowski Lagrangians is not a weak-field approximation but an exact Lagrangian for the electromagnetic field and the cosmological constant.
We show that this Lagrangian is not equivalent (dynamically) to the sum of the analogous ($\Lambda$CDM and Einstein-Maxwell) Lagrangians in the metric-affine/metric formulation (this inequivalence was already proved in Ref.~\cite{Proceed}).
This result is not surprising because what is simple (``minimal coupling'') in the metric formulation of gravity becomes complicated in the affine formulation, and what is simple in the affine formulation of gravity becomes complicated in the metric formulation~\cite{KW}.
Moreover, the purely affine Lagrangian for the Einstein-Klein-Gordon theory already shows that what is a sum in the metric picture is not a sum in the affine picture~\cite{Kij}.
In the same section we also show that the sum of the Eddington and Ferraris-Kijowski Lagrangians, like the Lagrangian in Sect.~\ref{secWeak}, is physical only for weak electromagnetic fields, on the order of the magnetic field in outer space of the Solar System.
In Sect.~\ref{secPA} we show the derivation of the purely affine equivalent of the Einstein-Maxwell-$\Lambda$ Lagrangian (this equivalent was found in Ref.~\cite{Proceed}).
This Lagrangian significantly differs from the sum of the affine Eddington and Ferraris-Kijowski Lagrangian, explaining why this sum does not describe physical systems.
We summarize the results in Sect.~\ref{secSum}.

\section{Field equations in purely affine gravity}
\label{secField}

The condition for a Lagrangian density to be covariant is that it must be a product of a scalar and the square root of the determinant of a covariant tensor of rank two, or a linear combination of such products~\cite{Edd,Schr}.
A general purely affine Lagrangian density $\textgoth{L}$ depends on the {\it affine connection} $\Gamma^{\,\,\rho}_{\mu\,\nu}$ (not restricted to be symmetric in the lower indices), the {\it curvature} tensor $R^\rho_{\phantom{\rho}\mu\sigma\nu}=\Gamma^{\,\,\rho}_{\mu\,\nu,\sigma}-\Gamma^{\,\,\rho}_{\mu\,\sigma,\nu}+\Gamma^{\,\,\kappa}_{\mu\,\nu}\Gamma^{\,\,\rho}_{\kappa\,\sigma}-\Gamma^{\,\,\kappa}_{\mu\,\sigma}\Gamma^{\,\,\rho}_{\kappa\,\nu}$, and their covariant derivatives (with respect to $\Gamma^{\,\,\rho}_{\mu\,\nu}$).
In order to be generally covariant, the Lagrangian density $\textgoth{L}$ may depend on $\Gamma^{\,\,\rho}_{\mu\,\nu}$ only through the covariant derivatives of tensors.
If we assume that $\textgoth{L}$ is of the first differential order with respect to the connection, as usually are Lagrangians in classical mechanics with respect to the configuration, and that derivatives of the connection appear in $\textgoth{L}$ only through the curvature $R^\rho_{\phantom{\rho}\mu\sigma\nu}$, then $\textgoth{L}$ is a function of $R^\rho_{\phantom{\rho}\mu\sigma\nu}$ and the {\it torsion} tensor $S^\rho_{\phantom{\rho}\mu\nu}=\Gamma^{\,\,\,\,\rho}_{[\mu\,\nu]}$: $\textgoth{L}=\textgoth{L}(S,R)$.
We assume that the dependence of $\textgoth{L}$ on the curvature is restricted to the {\it symmetric} part $P_{\mu\nu}=R_{(\mu\nu)}$ of the {\it Ricci} tensor $R_{\mu\nu}=R^\rho_{\phantom{\rho}\mu\rho\nu}$, as in general relativity: $\textgoth{L}=\textgoth{L}(S,P)$.\footnote{
For a general connection, there exist three independent traces of the curvature tensor: $P_{\mu\nu}$, the antisymmetric part of the Ricci tensor, $R_{[\mu\nu]}$, and the antisymmetric segmental~\cite{Weyl} (homothetic) curvature tensor (``second Ricci tensor''~\cite{Niko2}): $Q_{\mu\nu}=R^\rho_{\phantom{\rho}\rho\mu\nu}=\Gamma^{\,\,\rho}_{\rho\,\nu,\mu}-\Gamma^{\,\,\rho}_{\rho\,\mu,\nu}$, which has the form of a curl~\cite{Schr,Scho}.
}
We also assume that $\textgoth{L}$ depends on, in addition to the torsion tensor and the symmetrized Ricci tensor, a matter field $\phi$ and its covariant derivatives $\nabla\phi$: $\textgoth{L}=\textgoth{L}(S,P,\phi,\nabla\phi)$.
We denote this Lagrangian density as $\textgoth{L}(\Gamma,P,\phi,\partial\phi)$, bearing in mind that its dependence on the connection $\Gamma$ and ordinary derivatives $\partial\phi$ is not arbitrary, but such a Lagrangian is a covariant function of the torsion $S$ and $\nabla\phi$.
The variation of the corresponding action $I=\frac{1}{c}\int d^4x\textgoth{L}$ is given by
\begin{equation}
\delta I=\frac{1}{c}\int d^4x\Bigl(\frac{\partial\textgoth{L}}{\partial\Gamma^{\,\,\rho}_{\mu\,\nu}}\delta\Gamma^{\,\,\rho}_{\mu\,\nu}+\frac{\partial\textgoth{L}}{\partial P_{\mu\nu}}\delta P_{\mu\nu}+\frac{\partial\textgoth{L}}{\partial\phi}\delta\phi+\frac{\partial\textgoth{L}}{\partial\phi_{,\mu}}\delta(\phi_{,\mu})\Bigr).
\label{var1}
\end{equation}

The {\it fundamental} tensor density ${\sf g}^{\mu\nu}$ associated with a purely affine Lagrangian is obtained using~\cite{Kij,FK3a,FK3b,FK1,FK2a,FK2b}
\begin{equation}
{\sf g}^{\mu\nu}\equiv-2\kappa\frac{\partial\textgoth{L}}{\partial P_{\mu\nu}},
\label{met1}
\end{equation}
where $\kappa=\frac{8\pi G}{c^4}$ (for purely affine Lagrangians that depend on the symmetric part of the Ricci tensor, this definition is equivalent to that in Refs.~\cite{Edd,Schr,Niko2}: ${\sf g}^{\mu\nu}=-2\kappa\frac{\partial\textgoth{L}}{\partial R_{\mu\nu}}$).
This density introduces the {\it metric} structure in purely affine gravity by defining the symmetric contravariant metric tensor~\cite{Kij}:
\begin{equation}
g^{\mu\nu}\equiv\frac{{\sf g}^{\mu\nu}}{\sqrt{-\mbox{det}{\sf g}^{\rho\sigma}}}.
\label{met2}
\end{equation}
To make this definition meaningful, we must assume $\mbox{det}{\sf g}^{\mu\nu}\neq0$.
The physical signature requirement for $g_{\mu\nu}$ implies that we must take into account only those configurations with $\mbox{det}{\sf g}^{\mu\nu}<0$, which guarantees that $g_{\mu\nu}$ has the Lorentzian signature $(+,-,-,-)$ or $(-,+,+,+)$~\cite{KW}.
The symmetric covariant metric tensor $g_{\mu\nu}$ is related to the contravariant metric tensor by $g^{\mu\nu}g_{\rho\nu}=\delta^\mu_\rho$.\footnote{
If a purely affine Lagrangian depends on the full Ricci tensor, and the fundamental tensor density is defined as ${\sf g}^{\mu\nu}=-2\kappa\frac{\partial\textgoth{L}}{\partial R_{\mu\nu}}$, then ${\sf g}^{\mu\nu}$ is not symmetric.
This density introduces the metric structure in purely affine gravity by defining the symmetric contravariant metric tensor as $g^{\mu\nu}={\sf g}^{(\mu\nu)}/\sqrt{-\mbox{det}{\sf g}^{(\rho\sigma)}}$~\cite{Kur1,Kur2,Kur3}.
If we, instead, use Schr\"{o}dinger's definition $g^{\mu\nu}={\sf g}^{\mu\nu}/\sqrt{-\mbox{det}{\sf g}^{\rho\sigma}}$, the resulting field equations lead to the relation between the nonsymmetric metric and the affine connection in Einstein's generalized theory of gravitation~\cite{Schr,Schr1,Ein2}.
}
The tensors $g^{\mu\nu}$ and $g_{\mu\nu}$ are used for raising and lowering indices.

We also define the {\it hypermomentum} density conjugate to the affine connection~\cite{He1,He2,He3}:\footnote{
In Refs.~\cite{He1,He2,He3}, the hypermomentum density is defined as $\Pi_{\phantom{\mu}\rho\phantom{\nu}}^{\mu\phantom{\rho}\nu}=\frac{\partial\textgoth{L}}{\partial \Gamma^{\,\,\rho}_{\mu\,\nu}}$.
}
\begin{equation}
\Pi_{\phantom{\mu}\rho\phantom{\nu}}^{\mu\phantom{\rho}\nu}\equiv-2\kappa\frac{\partial\textgoth{L}}{\partial \Gamma^{\,\,\rho}_{\mu\,\nu}},
\label{con1}
\end{equation}
which has the same dimension as the connection.
Consequently, the variation of the action~(\ref{var1}) can be written as
\begin{equation}
\delta I=-\frac{1}{2\kappa c}\int d^4x(\Pi_{\phantom{\mu}\rho\phantom{\nu}}^{\mu\phantom{\rho}\nu}\delta\Gamma^{\,\,\rho}_{\mu\,\nu}+{\sf g}^{\mu\nu}\delta R_{\mu\nu})+\frac{1}{c}\int d^4x\Bigl(\frac{\partial\textgoth{L}}{\partial\phi}\delta\phi+\frac{\partial\textgoth{L}}{\partial\phi_{,\mu}}\delta(\phi_{,\mu})\Bigr).
\label{var2}
\end{equation}
If we assume that the field $\phi$ vanishes at the boundary of integration, then the field equation for $\phi$ is $\frac{\delta\textgoth{L}}{\delta\phi}=0$, where the variational derivative is defined as $\frac{\delta\textgoth{L}}{\delta\phi}=\frac{\partial\textgoth{L}}{\partial\phi}-\partial_\mu\Bigl(\frac{\partial\textgoth{L}}{\partial\phi_{,\mu}}\Bigr)$.
Accordingly, the variation~(\ref{var2}) takes the form:
\begin{equation}
\delta I=\frac{1}{c}\int d^4x\Bigl[-\frac{1}{2\kappa}\Bigl(\Pi_{\phantom{\mu}\rho\phantom{\nu}}^{\mu\phantom{\rho}\nu}\delta\Gamma^{\,\,\rho}_{\mu\,\nu}+{\sf g}^{\mu\nu}\delta P_{\mu\nu}\Bigl)+\frac{\delta\textgoth{L}}{\delta\phi}\delta\phi\Bigr].
\label{varphi}
\end{equation}

For a general affine connection, the variation of the Ricci tensor is given by the Palatini formula~\cite{Schr,Scho,He1}: $\delta R_{\mu\nu}=\delta\Gamma^{\,\,\rho}_{\mu\,\nu;\rho}-\delta\Gamma^{\,\,\rho}_{\mu\,\rho;\nu}-2S^\sigma_{\phantom{\sigma}\rho\nu}\delta\Gamma^{\,\,\rho}_{\mu\,\sigma}$, where the semicolon denotes the covariant differentiation with respect to $\Gamma^{\,\,\rho}_{\mu\,\nu}$.
Using the identity $\int d^4x({\sf V}^\mu)_{;\mu}=2\int d^4x S_\mu{\sf V}^\mu$, where ${\sf V}^\mu$ is an arbitrary contravariant vector density that vanishes at the boundary of the integration and $S_\mu=S^\nu_{\phantom{\nu}\mu\nu}$ is the torsion vector~\cite{Schr,Scho}, and applying the principle of least action $\delta S=0$ for arbitrary variations $\delta\Gamma^{\,\,\rho}_{\mu\,\nu}$, we obtain
\begin{equation}
{\sf g}^{\mu\nu}_{\phantom{\mu\nu};\rho}-{\sf g}^{\mu\sigma}_{\phantom{\mu\sigma};\sigma}\delta^\nu_\rho-2{\sf g}^{\mu\nu}S_\rho+2{\sf g}^{\mu\sigma}S_\sigma\delta^\nu_\rho+2{\sf g}^{\mu\sigma}S^\nu_{\phantom{\nu}\rho\sigma}=\Pi_{\phantom{\mu}\rho\phantom{\nu}}^{\mu\phantom{\rho}\nu}.
\label{field1}
\end{equation}
This equation is equivalent to
\begin{equation}
{\sf g}^{\mu\nu}_{\phantom{\mu\nu},\rho}+\,^\ast\Gamma^{\,\,\mu}_{\sigma\,\rho}{\sf g}^{\sigma\nu}+\,^\ast\Gamma^{\,\,\nu}_{\rho\,\sigma}{\sf g}^{\mu\sigma}-\,^\ast\Gamma^{\,\,\sigma}_{\sigma\,\rho}{\sf g}^{\mu\nu}=\Pi_{\phantom{\mu}\rho\phantom{\nu}}^{\mu\phantom{\rho}\nu}-\frac{1}{3}\Pi_{\phantom{\mu}\sigma\phantom{\sigma}}^{\mu\phantom{\sigma}\sigma}\delta^\nu_\rho,
\label{field2}
\end{equation}
where $^\ast\Gamma^{\,\,\rho}_{\mu\,\nu}=\Gamma^{\,\,\rho}_{\mu\,\nu}+\frac{2}{3}\delta^\rho_\mu S_\nu$~\cite{Schr,Schr1}.

Contracting the indices $\mu$ and $\rho$ in Eq.~(\ref{field1}) yields\footnote{
If the fundamental tensor density ${\sf g}^{\mu\nu}$ is defined as ${\sf g}^{\mu\nu}=-2\kappa\frac{\partial\textgoth{L}}{\partial R_{\mu\nu}}$ and thus, in general, is asymmetric, then Eq.~(\ref{cons2}) becomes: ${\sf g}^{[\nu\sigma]}_{\phantom{[\nu\sigma]},\sigma}+\frac{1}{2}\Pi_{\phantom{\sigma}\sigma\phantom{\nu}}^{\sigma\phantom{\sigma}\nu}=0$, which generalizes one of the field equations of Schr\"{o}dinger's nonsymmetric purely affine gravity~\cite{Schr,Schr1}.
}
\begin{equation}
\Pi_{\phantom{\sigma}\sigma\phantom{\nu}}^{\sigma\phantom{\sigma}\nu}=0,
\label{cons2}
\end{equation}
which is a constraint on how a purely affine Lagrangian depends on the connection.
This unphysical constraint is related to the fact that the gravitational part of this Lagrangian (proportional to ${\sf g}^{\mu\nu}P_{\mu\nu}$, see the next section) is invariant under projective transformations of the connection while the matter part, that can depend explicitly on the connection, is generally not invariant~\cite{He2,He3,San}.
We cannot assume that any form of matter will comply with this condition.
Therefore the field equations~(\ref{field1}) seem to be inconsistent.
To overcome this constraint we can restrict the torsion tensor to be traceless: $S_\mu=0$~\cite{San}.
Consequently, $^\ast\Gamma^{\,\,\rho}_{\mu\,\nu}=\Gamma^{\,\,\rho}_{\mu\,\nu}$.
This condition enters the Lagrangian density as a Lagrange multiplier term $-\frac{1}{2\kappa}{\sf B}^\mu S_\mu$, where the Lagrange multiplier ${\sf B}^\mu$ is a vector density.
Consequently, there is an extra term ${\sf B}^{[\mu}\delta^{\nu]}_\rho$ on the right-hand side of Eq.~(\ref{field1}) and Eq.~(\ref{cons2}) becomes $\frac{3}{2}{\sf B}^\nu=\Pi_{\phantom{\sigma}\sigma\phantom{\nu}}^{\sigma\phantom{\sigma}\nu}$.
Setting this equation to be satisfied identically removes the constraint~(\ref{cons2}) and brings Eq.~(\ref{field2}) into
\begin{equation}
{\sf g}^{\mu\nu}_{\phantom{\mu\nu},\rho}+\Gamma^{\,\,\mu}_{\sigma\,\rho}{\sf g}^{\sigma\nu}+\Gamma^{\,\,\nu}_{\rho\,\sigma}{\sf g}^{\mu\sigma}-\Gamma^{\,\,\sigma}_{\sigma\,\rho}{\sf g}^{\mu\nu}=\Pi_{\phantom{\mu}\rho\phantom{\nu}}^{\mu\phantom{\rho}\nu}-\frac{1}{3}\Pi_{\phantom{\mu}\sigma\phantom{\sigma}}^{\mu\phantom{\sigma}\sigma}\delta^\nu_\rho-\frac{1}{3}\Pi_{\phantom{\sigma}\sigma\phantom{\nu}}^{\sigma\phantom{\sigma}\nu}\delta^\mu_\rho.
\label{field3}
\end{equation}

Equation~(\ref{field3}) is an algebraic equation for $\Gamma^{\,\,\rho}_{\mu\,\nu}$ (with $S_\mu=0$) as a function of the metric tensor, its first derivatives and the density $\Pi_{\phantom{\mu}\rho\phantom{\nu}}^{\mu\phantom{\rho}\nu}$.
We seek its solution in the form:
\begin{equation}
\Gamma^{\,\,\rho}_{\mu\,\nu}=\{^{\,\,\rho}_{\mu\,\nu}\}_g+V^\rho_{\phantom{\rho}\mu\nu},
\label{sol1}
\end{equation}
where $\{^{\,\,\rho}_{\mu\,\nu}\}_g=\frac{1}{2}g^{\rho\sigma}(g_{\nu\sigma,\mu}+g_{\mu\sigma,\nu}-g_{\mu\nu,\sigma})$ is the Christoffel connection of the metric tensor $g_{\mu\nu}$.
Consequently, the Ricci tensor of the affine connection $\Gamma^{\,\,\rho}_{\mu\,\nu}$ is given by~\cite{Scho}
\begin{equation}
R_{\mu\nu}(\Gamma)=R_{\mu\nu}(g)+V^\rho_{\phantom{\rho}\mu\nu:\rho}-V^\rho_{\phantom{\rho}\mu\rho:\nu}+V^\sigma_{\phantom{\sigma}\mu\nu}V^\rho_{\phantom{\rho}\sigma\rho}-V^\sigma_{\phantom{\sigma}\mu\rho}V^\rho_{\phantom{\rho}\sigma\nu},
\label{sol2}
\end{equation}
where $R_{\mu\nu}(g)$ is the Riemannian Ricci tensor of the metric tensor $g_{\mu\nu}$ and the colon denotes the covariant differentiation with respect to $\{^{\,\,\rho}_{\mu\,\nu}\}_g$.
Substituting Eq.~(\ref{sol1}) to Eq.~(\ref{field3}) gives
\begin{equation}
V^\mu_{\phantom{\mu}\sigma\rho}{\sf g}^{\sigma\nu}+V^\nu_{\phantom{\nu}\rho\sigma}{\sf g}^{\mu\sigma}-V^\sigma_{\phantom{\sigma}\sigma\rho}{\sf g}^{\mu\nu}=\Pi_{\phantom{\mu}\rho\phantom{\nu}}^{\mu\phantom{\rho}\nu}-\frac{1}{3}\Pi_{\phantom{\mu}\sigma\phantom{\sigma}}^{\mu\phantom{\sigma}\sigma}\delta^\nu_\rho-\frac{1}{3}\Pi_{\phantom{\sigma}\sigma\phantom{\nu}}^{\sigma\phantom{\sigma}\nu}\delta^\mu_\rho,
\label{sol3}
\end{equation}
which is a linear relation between $V^\rho_{\phantom{\rho}\mu\nu}$ and $\Pi_{\phantom{\mu}\rho\phantom{\nu}}^{\mu\phantom{\rho}\nu}$ and can be solved~\cite{unif1,unif2}.

If a purely affine Lagrangian does not depend explicitly on the connection (such Lagrangians are studied later in this paper) then $\Pi_{\phantom{\mu}\rho\phantom{\nu}}^{\mu\phantom{\rho}\nu}=0$.
In this case, we do not need to introduce the condition $S_\mu=0$ (or any constraint on four degrees of freedom of the connection) and Eq.~(\ref{field2}) becomes
\begin{equation}
{\sf g}^{\mu\nu}_{\phantom{\mu\nu},\rho}+\,^\ast\Gamma^{\,\,\mu}_{\sigma\,\rho}{\sf g}^{\sigma\nu}+\,^\ast\Gamma^{\,\,\nu}_{\rho\,\sigma}{\sf g}^{\mu\sigma}-\,^\ast\Gamma^{\,\,\sigma}_{\sigma\,\rho}{\sf g}^{\mu\nu}=0.
\label{field4}
\end{equation}
The tensor $P_{\mu\nu}$ is invariant under a projective transformation $\Gamma^{\,\,\rho}_{\mu\,\nu}\rightarrow\Gamma^{\,\,\rho}_{\mu\,\nu}+\delta^\rho_\mu W_\nu$.
We can use this transformation, with $W_\mu=\frac{2}{3}S_\mu$, to bring the torsion vector $S_\mu$ to zero and make $^\ast\Gamma^{\,\,\rho}_{\mu\,\nu}=\Gamma^{\,\,\rho}_{\mu\,\nu}$.
From Eq.~(\ref{field4}) it follows that the affine connection is the Christoffel connection of the metric tensor:
\begin{equation}
\Gamma^{\,\,\rho}_{\mu\,\nu}=\{^{\,\,\rho}_{\mu\,\nu}\}_g,
\label{Chr}
\end{equation}
which is the special case of Eq.~(\ref{sol1}) with $V^\rho_{\phantom{\rho}\mu\nu}=0$.

The theory based on a general Lagrangian density $\textgoth{L}(S,P,\phi,\nabla\phi)$ without any constraints on the affine connection does not determine the connection uniquely because the tensor $P_{\mu\nu}$ is invariant under projective transformations of the connection, $\Gamma^{\,\,\rho}_{\mu\,\nu}\rightarrow\Gamma^{\,\,\rho}_{\mu\,\nu}+\delta^\rho_\mu V_\nu$, where $V_\nu$ is a vector function of the coordinates.
Therefore at least four degrees of freedom must be constrained to make such a theory consistent from a physical point of view~\cite{He2,He3}.
The condition $S_\mu=0$ is not the only way to impose such a constraint; other possibilities include vanishing of the Weyl vector $W_\nu=\frac{1}{2}(\Gamma^{\,\,\rho}_{\rho\,\nu}-\{^{\,\,\rho}_{\rho\,\nu}\}_g)=0$~\cite{He2,He3} or adding the dependence on the segmental curvature tensor, which is not projectively invariant, to the Lagrangian~\cite{He2,He3,unif1,unif2}.\footnote{
If a purely affine Lagrangian density depends also on the segmental curvature tensor $Q_{\mu\nu}$, then we can define the antisymmetric tensor density ${\sf h}^{\mu\nu}\equiv-2\kappa\frac{\delta{\cal L}_Q}{\delta Q_{\mu\nu}}$.
In this case, Eq.~(\ref{field2}) becomes:
\begin{equation}
{\sf g}^{\mu\nu}_{\phantom{\mu\nu},\rho}+\,^\ast\Gamma^{\,\,\mu}_{\sigma\,\rho}{\sf g}^{\sigma\nu}+\,^\ast\Gamma^{\,\,\nu}_{\rho\,\sigma}{\sf g}^{\mu\sigma}-\,^\ast\Gamma^{\,\,\sigma}_{\sigma\,\rho}{\sf g}^{\mu\nu}=\Pi_{\phantom{\mu}\rho\phantom{\nu}}^{\mu\phantom{\rho}\nu}-\frac{1}{3}\Pi_{\phantom{\mu}\sigma\phantom{\sigma}}^{\mu\phantom{\sigma}\sigma}\delta^\nu_\rho+2{\sf h}^{\nu\sigma}_{\phantom{\nu\sigma},\sigma}\delta^\mu_\rho-\frac{2}{3}{\sf h}^{\mu\sigma}_{\phantom{\mu\sigma},\sigma}\delta^\nu_\rho,
\label{fieldFK}
\end{equation}
and the unphysical algebraic Eq.~(\ref{cons2}) turns into a dynamical field equation for ${\sf h}^{\mu\nu}$:
\begin{equation}
{\sf h}^{\sigma\nu}_{\phantom{\sigma\nu},\sigma}=\frac{1}{8}\Pi_{\phantom{\sigma}\sigma\phantom{\nu}}^{\sigma\phantom{\sigma}\nu},
\label{cons1}
\end{equation}
which looks like the second pair of the Maxwell equations with the source represented by the trace of the hypermomentum density~\cite{Niko3}.
}

If we assume that the affine connection in the purely affine variational principle is symmetric, as in the original formulation of purely affine gravity~\cite{Kij,FK3a,FK3b}, then instead of Eq.~(\ref{field1}) we find~\cite{KW}:
\begin{equation}
{\sf g}^{\mu\nu}_{\phantom{\mu\nu};\rho}-{\sf g}^{\sigma(\mu}_{\phantom{\sigma(\mu};\sigma}\delta^{\nu)}_\rho=\Pi_{\phantom{\mu}\rho\phantom{\nu}}^{\mu\phantom{\rho}\nu}.
\label{fieldsym}
\end{equation}
The hypermomentum density $\Pi_{\phantom{\mu}\rho\phantom{\nu}}^{\mu\phantom{\rho}\nu}$ is, because of the definition~(\ref{con1}), symmetric in the upper indices.
Contracting the indices $\mu$ and $\rho$ in Eq.~(\ref{fieldsym}) does not lead to any algebraic constraint on $\Pi_{\phantom{\mu}\rho\phantom{\nu}}^{\mu\phantom{\rho}\nu}$.
Therefore, for purely affine Lagrangians that depend only on the connection and the symmetric part of the Ricci tensor, the relation $S_\mu=0$ appears either as a remedy for the unphysical constraint on the hypermomentum density, or results simply from using a symmetric affine connection as a dynamical variable.
In this paper we study purely affine Lagrangians that depend on the connection only via the symmetrized Ricci tensor: $\textgoth{L}=\textgoth{L}(P,\phi,\partial\phi)$, so $S_\mu$ decouples from the Einstein equations and can be brought to zero by a projective transformation, leading to Eq.~(\ref{Chr}).

\section{Equivalence of affine, metric-affine and metric pictures}
\label{secCor}

If we apply to a purely affine Lagrangian $\textgoth{L}(\Gamma,P,\phi,\partial\phi)$ the Legendre transformation with respect to $P_{\mu\nu}$~\cite{Kij,FK3a,FK3b}, defining the Hamiltonian density $\textgoth{H}$ (or rather the Routhian density~\cite{LL1}, since we can also apply the Legendre transformation with respect to $\Gamma^{\,\,\rho}_{\mu\,\nu}$):
\begin{equation}
\textgoth{H}=\textgoth{L}-\frac{\partial\textgoth{L}}{\partial P_{\mu\nu}}P_{\mu\nu}=\textgoth{L}+\frac{1}{2\kappa}{\sf g}^{\mu\nu}P_{\mu\nu},
\label{Leg1}
\end{equation}
we find for the differential $d\textgoth{H}$:
\begin{equation}
d\textgoth{H}=\frac{\partial\textgoth{L}}{\partial \Gamma^{\,\,\rho}_{\mu\,\nu}}d\Gamma^{\,\,\rho}_{\mu\,\nu}+\frac{1}{2\kappa}P_{\mu\nu}d{\sf g}^{\mu\nu}+\frac{\partial\textgoth{L}}{\partial\phi}d\phi+\frac{\partial\textgoth{L}}{\partial\phi_{,\mu}}d\phi_{,\mu}.
\label{Leg2}
\end{equation}
Accordingly, the Hamiltonian density $\textgoth{H}$ is a covariant function of $\Gamma^{\,\,\rho}_{\mu\,\nu}$, ${\sf g}^{\mu\nu}$, $\phi$ and $\partial\phi$: $\textgoth{H}=\textgoth{H}(\Gamma,{\sf g},\phi,\partial\phi)=\textgoth{H}(S,{\sf g},\phi,\nabla\phi)$, and the action variation~(\ref{varphi}) takes the form:
\begin{eqnarray}
& & \delta I=\frac{1}{c}\delta\int d^4x\biggl(\textgoth{H}(\Gamma,{\sf g},\phi,\partial\phi)-\frac{1}{2\kappa}{\sf g}^{\mu\nu}P_{\mu\nu}\biggr) \nonumber \\
& & =\frac{1}{c}\int d^4x\biggl(\frac{\partial\textgoth{H}}{\partial\Gamma^{\,\,\rho}_{\mu\,\nu}}\delta\Gamma^{\,\,\rho}_{\mu\,\nu}+\frac{\partial\textgoth{H}}{\partial{\sf g}^{\mu\nu}}\delta{\sf g}^{\mu\nu}-\frac{1}{2\kappa}{\sf g}^{\mu\nu}\delta P_{\mu\nu}-\frac{1}{2\kappa}P_{\mu\nu}\delta{\sf g}^{\mu\nu}+\frac{\delta\textgoth{H}}{\delta\phi}\delta\phi\biggr). \nonumber \\
& & 
\label{var4}
\end{eqnarray}
The variation with respect to ${\sf g}^{\mu\nu}$ yields the first Hamilton equation~\cite{Kij,FK3a,FK3b}:
\begin{equation}
P_{\mu\nu}=2\kappa\frac{\partial\textgoth{H}}{\partial {\sf g}^{\mu\nu}}.
\label{Ham1}
\end{equation}
The variation with respect to $P_{\mu\nu}$ can be transformed to the variation with respect to $\Gamma^{\,\,\rho}_{\mu\,\nu}$ by means of the Palatini formula, giving the second Hamilton equation equivalent to the field equations~(\ref{field1}).
The field equation for $\phi$: $\frac{\delta\textgoth{H}}{\delta\phi}=0$, is equivalent to $\frac{\delta\textgoth{L}}{\delta\phi}=0$, so $\textgoth{H}(\Gamma,{\sf g},\phi,\partial\phi)$ plays the role of the matter Lagrangian for the field $\phi$~\cite{Kij,FK3a,FK3b,KW}.

The analogous transformation in classical mechanics goes from a Lagrangian $L(q^i,\dot{q}^i)$ to a Hamiltonian $H(q^i,p^i)=p^j\dot{q}^j-L(q^i,\dot{q}^i)$ with $p^i=\frac{\partial{L}}{\partial\dot{q}^i}$, where the symmetrized Ricci tensor $P_{\mu\nu}$ corresponds to {\em generalized velocities} $\dot{q}^i$ and the fundamental density ${\sf g}^{\mu\nu}$ to {\em canonical momenta} $p^i$~\cite{Kij,FK3a,FK3b}.
Accordingly, the affine connection $\Gamma^{\,\,\rho}_{\mu\,\nu}$ plays the role of the {\em configuration} $q^i$ and the hypermomentum density $\Pi_{\phantom{\mu}\rho\phantom{\nu}}^{\mu\phantom{\rho}\nu}$ corresponds to {\em generalized forces} $f^i=\frac{\partial{L}}{\partial q^i}$~\cite{Kij}.
The field equations~(\ref{field2}) correspond to the Lagrange equations $\frac{\partial L}{\partial q^i}=\frac{d}{dt}\frac{\partial L}{\partial\dot{q}^i}$ which result from Hamilton's principle $\delta\int L(q^i,\dot{q}^i)dt=0$ for arbitrary variations $\delta q^i$ vanishing at the boundaries of the configuration.
The Hamilton equations result from the same principle written as $\delta\int(p^j\dot{q}^j-H(q^i,p^i))dt=0$ for arbitrary variations $\delta q^i$ and $\delta p^i$~\cite{LL1}.
The field equations~(\ref{field1}) correspond to the second Hamilton equation, $\dot{p}^i=-\frac{\partial H}{\partial q^i}$, and Eq.~(\ref{Ham1}) to the first Hamilton equation, $\dot{q}^i=\frac{\partial H}{\partial p^i}$.

Equation~(\ref{Ham1}) gives the metric-affine Einstein equations:
\begin{equation}
\sqrt{-g}\Bigl(P_{\mu\nu}-\frac{1}{2}Pg_{\mu\nu}\Bigr)=2\kappa\frac{\partial\textgoth{H}}{\partial g^{\mu\nu}},
\label{Ham2}
\end{equation}
where $P=P_{\mu\nu}g^{\mu\nu}$ and $g=\mbox{det}g_{\mu\nu}=\mbox{det}{\sf g}^{\mu\nu}$.
Since $\textgoth{H}(\Gamma,{\sf g},\phi,\partial\phi)$ plays the role of the matter Lagrangian for the field $\phi$, the right-hand side of Eq.~(\ref{Ham2}) is the symmetric dynamical energy-momentum tensor $T_{\mu\nu}$ for $\phi$ in the metric-affine formulation:
\begin{equation}
\sqrt{-g}T_{\mu\nu}=2\frac{\partial\textgoth{H}}{\partial g^{\mu\nu}}.
\label{EMTen}
\end{equation}
The derivative of $\textgoth{H}(\Gamma,{\sf g},\phi,\partial\phi)$ with respect to the connection is related to the hypermomentum density:
\begin{equation}
\Pi_{\phantom{\mu}\rho\phantom{\nu}}^{\mu\phantom{\rho}\nu}=-2\kappa\frac{\partial\textgoth{H}}{\partial \Gamma^{\,\,\rho}_{\mu\,\nu}}.
\label{hypDen}
\end{equation}
From the first line in Eq.~(\ref{var4}) it follows that $-\frac{1}{2\kappa}P\sqrt{-g}$ is the metric-affine Lagrangian density for the gravitational field, in agreement with the general-relativistic form.
According to Eq.~(\ref{Leg1}), the numerical value of the affine Lagrangian $\textgoth{L}$ coincides with the metric Lagrangian for gravity and matter, with the metric tensor eliminated by means of Eq.~(\ref{met1}).

The transition from the purely affine formalism to the metric-affine one shows that the gravitational Lagrangian density $\mathcal{L}_g$ is a {\em Legendre term} corresponding to $p^j\dot{q}^j$ in classical mechanics~\cite{Kij}.
Therefore the purely affine and metric-affine formulation of gravitation are dynamically equivalent, if $\textgoth{L}$ depends on the affine connection and the symmetric part of the Ricci tensor~\cite{FK3a,FK3b}.
The field equations in one formulation become the definitions of canonically conjugate quantities in another, and vice versa~\cite{FK3a,FK3b}.
Equations~(\ref{Ham2}) and~(\ref{EMTen}) with symmetrized Eq.~(\ref{sol2}) give
\begin{eqnarray}
& & R_{\mu\nu}(g)=\kappa T_{\mu\nu}-\frac{\kappa}{2}T_{\rho\sigma}g^{\rho\sigma}g_{\mu\nu}-V^\rho_{\phantom{\rho}(\mu\nu):\rho}+V^\rho_{\phantom{\rho}(\mu|\rho:|\nu)} \nonumber \\
& & -V^\sigma_{\phantom{\sigma}(\mu\nu)}V^\rho_{\phantom{\rho}\sigma\rho}+V^\sigma_{\phantom{\sigma}(\mu|\rho}V^\rho_{\phantom{\rho}\sigma|\nu)},
\label{EMT}
\end{eqnarray}
Combining Eqs.~(\ref{sol3}) and~(\ref{EMT}) yields a relation between the Riemannian Ricci tensor $R_{\mu\nu}(g)$ and the hypermomentum density $\Pi_{\phantom{\mu}\rho\phantom{\nu}}^{\mu\phantom{\rho}\nu}$, i.e. the purely metric Einstein equations.
The tensor $T_{\mu\nu}$ represents the matter part that is generated by the metric tensor, e.g., the electromagnetic field.
The terms in Eq.~(\ref{EMT}) that contain $V^\rho_{\phantom{\rho}\mu\nu}$ form the tensor that we denote as $\kappa(\Theta_{\mu\nu}-\frac{1}{2}\Theta_{\rho\sigma}g^{\rho\sigma}g_{\mu\nu})$.
The symmetric tensor $\Theta_{\mu\nu}$, corresponding to the matter part that is generated by the connection, contains contributions from torsion and is quadratic in the density $\Pi_{\phantom{\mu}\rho\phantom{\nu}}^{\mu\phantom{\rho}\nu}$.
The Bianchi identity for the tensor $R_{\mu\nu}(g)$ yields the covariant conservation of the total energy-momentum tensor: $(T^{\mu\nu}+\Theta^{\mu\nu})_{:\nu}=0$.

The purely metric (standard general-relativistic) formulation is dynamically equivalent to the purely affine and metric-affine formulations, which can be shown by applying to $\textgoth{H}(\Gamma^{\,\,\rho}_{\mu\,\nu},{\sf g}^{\mu\nu},\phi,\phi_{,\mu})$ the Legendre transformation with respect to $\Gamma^{\,\,\rho}_{\mu\,\nu}$ or by demonstrating that the second-order derivatives of the metric tensor in the energy-momentum tensor for matter in the purely metric formulation of gravity, together with the Riemannian Ricci tensor, sum up to the Ricci tensor of a new connection~\cite{FK3a,FK3b,KW}.
In the following sections we will consider purely affine Lagrangians that do not depend explicitly on the affine connection and for which the metric-affine and purely metric formulation are equivalent straightforwardly, so the equivalence of the purely affine and purely metric picture follows directly from the equivalence of the purely affine and metric-affine picture that is associated with the Legendre transformation with respect to $\Gamma^{\,\,\rho}_{\mu\,\nu}$.

We note that the metric-affine Lagrangian density for the gravitational field $\mathcal{L}_g$ automatically turns out to be linear in the curvature tensor~\cite{FK3a,FK3b}.
The purely metric Lagrangian density for the gravitational field turns out to be linear in the curvature tensor as well since $P$ is a linear function of $R_{\mu\nu}(g)g^{\mu\nu}$.
Thus metric-affine and metric Lagrangians for the gravitational field that are nonlinear with respect to curvature {\em cannot} be derived from a purely affine Lagrangian $\textgoth{L}(\Gamma^{\,\,\rho}_{\mu\,\nu},P_{\mu\nu},\phi,\phi_{,\mu})=\textgoth{L}(S^\rho_{\phantom{\rho}\mu\nu},P_{\mu\nu},\phi,\phi_{;\mu})$.

\section{Eddington Lagrangian}
\label{secEdd}

The simplest purely affine Lagrangian density $\textgoth{L}=\textgoth{L}(P_{\mu\nu})$ was introduced by Eddington~\cite{Edd,FK1}:
\begin{equation}
\textgoth{L}_{\textrm{\scriptsize{Edd}}}=\frac{1}{\kappa\Lambda}\sqrt{-\mbox{det}P_{\mu\nu}}.
\label{Lagr1}
\end{equation}
To make this Lagrangian density meaningful, we assume $\mbox{det}P_{\mu\nu}<0$, that is, the symmetrized Ricci tensor $P_{\mu\nu}$ has the Lorentzian signature.
The Eddington Lagrangian does not depend explicitly on the affine connection, which is analogous in classical mechanics to free Lagrangians that depend only on generalized velocities: $L=L(\dot{q}^i)$.
Accordingly, the Lagrangian density~(\ref{Lagr1}) describes a free gravitational field.

Substituting Eq.~(\ref{Lagr1}) into Eq.~(\ref{met1}) yields~\cite{Schr}
\begin{equation}
{\sf g}^{\mu\nu}=-\frac{1}{\Lambda}\sqrt{-\mbox{det}P_{\rho\sigma}}P^{\mu\nu},
\label{cosm1}
\end{equation}
where the symmetric tensor $P^{\mu\nu}$ is reciprocal to the symmetrized Ricci tensor $P_{\mu\nu}$: $P^{\mu\nu}P_{\rho\nu}=\delta^\mu_\rho$.
Equation~(\ref{cosm1}) is equivalent to
\begin{equation}
P_{\mu\nu}=-\Lambda g_{\mu\nu}.
\label{cosm2}
\end{equation} 
Since the Lagrangian density~(\ref{Lagr1}) does not depend explicitly on the connection, the field equations are given by Eq.~(\ref{Chr}).
As a result, Eq.~(\ref{cosm2}) becomes
\begin{equation}
R_{\mu\nu}(g)=-\Lambda g_{\mu\nu},
\label{cosm3}
\end{equation}
which has the form of the Einstein field equations of general relativity with the cosmological constant $\Lambda$~\cite{Edd,Schr}.
This equivalence can also be shown by using Eqs.~(\ref{Leg1}) and~(\ref{cosm2}).
If $\textgoth{L}=\textgoth{L}_{\textrm{\scriptsize{Edd}}}$ then $\textgoth{H}=\textgoth{H}_\Lambda$, where
\begin{equation}
\textgoth{H}_\Lambda=-\frac{\Lambda}{\kappa}\sqrt{-g},
\label{cosm4}
\end{equation}
which is the same as the Einstein (or ($\Lambda$CDM) metric-affine Lagrangian density for the cosmological constant.
Applying to the Eddington Hamiltonian density $\textgoth{H}_\Lambda$ the Legendre transformation with respect to the connection $\Gamma^{\,\,\rho}_{\mu\,\nu}$ does not do anything since $\textgoth{H}_\Lambda$ does not depend on $\Gamma^{\,\,\rho}_{\mu\,\nu}$.
Thus the purely metric Lagrangian density for the cosmological constant equals $\textgoth{H}_\Lambda$.

We note that the purely affine formulation of general relativity is not completely equivalent to the metric-affine and metric formulation because of one feature: it is {\em impossible} to find a purely affine Lagrangian that produces the Einstein equations in vacuum $R_{\mu\nu}=0$.
In fact, from the definitions~(\ref{met1}) and~(\ref{met2}) we obtain the relation between the Ricci tensor and the contravariant metric tensor.
A free gravitational field depends only on the Ricci tensor thus this relation has the form $g_{\mu\nu}=f(R_{\alpha\beta})$ or, inversely, $R_{\mu\nu}(\Gamma)=f^{-1}(g_{\alpha\beta})$.
Consequently, the definition of the metric density as the Hamiltonian derivative of the Lagrangian density with respect to the Ricci tensor requires that the latter is algebraically related to the metric tensor.

The simplest purely affine Lagrangian, of Eddington, yields this relation in the form of Eq.~(\ref{cosm2}), i.e. automatically generates a cosmological constant (without specifying its sign).
This mechanism is supported by cosmological observations reporting that the universe is currently accelerating~\cite{acc1,acc2} in accordance with general relativity with a constant, positive ($\Lambda>0$) cosmological term (the $\Lambda$CDM model)~\cite{const}.\footnote{
The value of $\Lambda$ is on the order of $10^{-52}\textrm{m}^{-2}$.
}
Therefore this restriction of the purely affine formulation of general relativity turns out to be its advantage.
We also note that another restriction of the purely affine formulation of general relativity, imposing that the metric-affine and metric Lagrangians must be linear in the Ricci scalar~\cite{Niko4} and thus excluding modified theories of gravity such as $f(R)$ models, turns out to be supported by cosmological and Solar System observations~\cite{linear}.

\section{Ferraris-Kijowski Lagrangian}
\label{secFK}

The purely affine Lagrangian density of Ferraris and Kijowski~\cite{FK1}:
\begin{equation}
\textgoth{L}_{\textrm{\scriptsize{FK}}}=-\frac{1}{4}\sqrt{-\mbox{det}P_{\mu\nu}}F_{\alpha\beta}F_{\rho\sigma}P^{\alpha\rho}P^{\beta\sigma},
\label{Lagr2}
\end{equation}
where $\mbox{det}P_{\mu\nu}<0$, has the form of the metric-affine (or metric, since the connection does not appear explicitly) Maxwell Lagrangian density for the electromagnetic field $F_{\mu\nu}$:\footnote{
The electromagnetic field tensor is defined as $F_{\mu\nu}=A_{\nu,\mu}-A_{\mu,\nu}=A_{\nu:\mu}-A_{\mu:\nu}$ so the electromagnetic potential $A_\mu$ does not couple to the torsion~\cite{HehlEM}.
}
\begin{equation}
\textgoth{H}_{\textrm{\scriptsize{Max}}}=-\frac{1}{4}\sqrt{-g}F_{\alpha\beta}F_{\rho\sigma}g^{\alpha\rho}g^{\beta\sigma},
\label{Lagr4}
\end{equation}
in which the covariant metric tensor is replaced by the symmetrized Ricci tensor $P_{\mu\nu}$ and the contravariant metric tensor by the tensor $P^{\mu\nu}$ reciprocal to $P_{\mu\nu}$.
Substituting Eq.~(\ref{Lagr2}) to Eq.~(\ref{met1}) gives (in purely affine picture)
\begin{equation}
{\sf g}^{\mu\nu}=\kappa\sqrt{-\mbox{det}P_{\rho\sigma}}P^{\beta\sigma}F_{\alpha\beta}F_{\rho\sigma}\biggl(\frac{1}{4}P^{\mu\nu}P^{\alpha\rho}-P^{\mu\alpha}P^{\nu\rho}\biggr).
\label{extra1}
\end{equation}
From Eqs.~(\ref{Ham1}) and~(\ref{Lagr4}) it follows that (in metric-affine/metric picture)
\begin{equation}
P_{\mu\nu}-\frac{1}{2}Pg_{\mu\nu}=\kappa\biggl(\frac{1}{4}F_{\alpha\beta}F_{\rho\sigma}g^{\alpha\rho}g^{\beta\sigma}g_{\mu\nu}-F_{\mu\alpha}F_{\nu\beta}g^{\alpha\beta}\biggr),
\label{EinMax1}
\end{equation}
which yields $P=0$.
Consequently, Eq.~(\ref{Leg1}) reads $\textgoth{L}_{\textrm{\scriptsize{Max}}}=\textgoth{H}_{\textrm{\scriptsize{Max}}}$, where $\textgoth{L}_{\textrm{\scriptsize{Max}}}$ is the purely affine Lagrangian density that is dynamically equivalent to the Maxwell Lagrangian density~(\ref{Lagr4}).
Similarly, $\textgoth{H}_{\textrm{\scriptsize{FK}}}=\textgoth{L}_{\textrm{\scriptsize{FK}}}$, where $\textgoth{H}_{\textrm{\scriptsize{FK}}}$ is the metric-affine density that is dynamically equivalent to the Ferraris-Kijowski Lagrangian density~(\ref{Lagr4}).

The Lagrangian density~(\ref{Lagr2}) is dynamically equivalent to the Lagrangian density~(\ref{Lagr4}), i.e. $\textgoth{H}=\textgoth{H}_{\textrm{\scriptsize{Max}}}$ is equivalent to $\textgoth{L}=\textgoth{L}_{\textrm{\scriptsize{FK}}}$~\cite{FK1}, which means that Eqs.~(\ref{extra1}) and~(\ref{EinMax1}) are equivalent.
Or, in other words, $\textgoth{H}_{\textrm{\scriptsize{FK}}}=\textgoth{H}_{\textrm{\scriptsize{Max}}}$ and $\textgoth{L}_{\textrm{\scriptsize{Max}}}=\textgoth{L}_{\textrm{\scriptsize{FK}}}$.
To see that the Lagrangian density~(\ref{Lagr2}) indeed represents the Maxwell electrodynamics, it is sufficient to choose the frame of reference in which $g_{\mu\nu}$ is galilean:
\begin{equation}
g_{\mu\nu}=\left( \begin{array}{cccc}
1 & 0 & 0 & 0 \\
0 & -1 & 0 & 0 \\
0 & 0 & -1 & 0 \\
0 & 0 & 0 & -1 \end{array} \right),
\label{Gal}
\end{equation}
and the electric ${\bf E}$ and magnetic ${\bf B}$ vectors are parallel (where the $x$ axis is taken along the direction of the field)~\cite{FK1}:
\begin{equation}
F_{\mu\nu}=\left( \begin{array}{cccc}
0 & E & 0 & 0 \\
-E & 0 & 0 & 0 \\
0 & 0 & 0 & -B \\
0 & 0 & B & 0 \end{array} \right).
\label{EM}
\end{equation}
In this case, Eq.~(\ref{EinMax1}) yields a diagonal tensor,
\begin{equation}
P_{\mu\nu}=\frac{\kappa}{2}(E^2+B^2)\left( \begin{array}{cccc}
1 & 0 & 0 & 0 \\
0 & -1 & 0 & 0 \\
0 & 0 & 1 & 0 \\
0 & 0 & 0 & 1 \end{array} \right),
\label{REM1}
\end{equation}
which gives the desired formula:\footnote{
In the chosen frame of reference, both sides of Eq.~(\ref{REM2}) are equal to the matrix
\begin{equation}
\left( \begin{array}{cccc}
0 & -E & 0 & 0 \\
E & 0 & 0 & 0 \\
0 & 0 & 0 & -B \\
0 & 0 & B & 0 \end{array} \right).
\label{REM3}
\end{equation}}
\begin{equation}
\sqrt{-\mbox{det}P_{\mu\nu}}F_{\alpha\beta}P^{\alpha\rho}P^{\beta\sigma}=\sqrt{-g}F_{\alpha\beta}g^{\alpha\rho}g^{\beta\sigma}.
\label{REM2}
\end{equation}
This expression is of tensorial character, hence it is valid in any frame of reference.\footnote{Taking the determinant of both sides of Eq.~(\ref{REM2}) gives the identity.}
Therefore, the Lagrangian densities~(\ref{Lagr2}) and~(\ref{Lagr4}) are equivalent~\cite{FK1}.

The sourceless Maxwell equations in purely affine gravity follow from varying the action corresponding to the Lagrangian density~(\ref{Lagr2}) with respect to $A_\mu$ and are given by
\begin{equation}
\biggl(\sqrt{-\mbox{det}P_{\alpha\beta}}F_{\rho\sigma}P^{\mu\rho}P^{\nu\sigma}\biggr)_{,\mu}=0.
\label{Max1}
\end{equation}
Using Eq.~(\ref{REM2}), Eq.~(\ref{Max1}) becomes
\begin{equation}
(\sqrt{-g}F^{\mu\nu})_{,\mu}=\sqrt{-g}F^{\mu\nu}_{\phantom{\mu\nu}:\mu}=0,
\label{Max2}
\end{equation}
as in the metric formulation, and is equivalent, due to Eq.~(\ref{Chr}), to
\begin{equation}
F^{\mu\nu}_{\phantom{\mu\nu};\mu}=0.
\label{Max3}
\end{equation}

The tensor $P_{\mu\nu}$ was brought to diagonal form by transforming to a reference system in which the vectors ${\bf E}$ and ${\bf B}$ (at the given point in spacetime) are parallel to one another.
Such a transformation is always possible except when ${\bf E}$ and ${\bf B}$ are mutually perpendicular and equal in magnitude~\cite{LL2}.\footnote{
The reduction of the tensor $P_{\mu\nu}$ to principal axes may be impossible because the spacetime is pseudo-euclidean.
}
But if the vectors ${\bf E}$ and ${\bf B}$ are mutually perpendicular and equal in magnitude, as in the case of a plane electromagnetic wave, the tensor $P_{\mu\nu}$ cannot be brought to diagonal form, as in~\cite{FK1}.
If the $x$ axis is taken along the direction of ${\bf E}$ and the $y$ axis along ${\bf B}$:
\begin{equation}
F_{\mu\nu}=\left( \begin{array}{cccc}
0 & E & 0 & 0 \\
-E & 0 & 0 & B \\
0 & 0 & 0 & 0 \\
0 & -B & 0 & 0 \end{array} \right),
\label{REM4}
\end{equation}
the tensor $P_{\mu\nu}$ becomes
\begin{equation}
P_{\mu\nu}=\kappa\left( \begin{array}{cccc}
\frac{E^2+B^2}{2} & 0 & 0 & -EB \\
0 & \frac{B^2-E^2}{2} & 0 & 0 \\
0 & 0 & \frac{E^2-B^2}{2} & 0 \\
-EB & 0 & 0 & \frac{E^2+B^2}{2} \end{array} \right).
\label{REM5}
\end{equation}
If $E=B$, the determinant $\mbox{det}P_{\mu\nu}$ vanishes and we cannot construct the reciprocal tensor $P^{\mu\nu}$.
However, we can regard the singular case $E=B$ as the limit $E\rightarrow B$, for which $\sqrt{-\mbox{det}P_{\mu\nu}}F_{\alpha\beta}P^{\alpha\rho}P^{\beta\sigma}$ is well-defined, obtaining
\begin{equation}
\sqrt{-\mbox{det}P_{\mu\nu}}F_{\alpha\beta}P^{\alpha\rho}P^{\beta\sigma}=\left( \begin{array}{cccc}
0 & -E & 0 & 0 \\
E & 0 & 0 & B \\
0 & 0 & 0 & 0 \\
0 & -B & 0 & 0 \end{array} \right).
\label{REM6}
\end{equation}
This expression is equal to $\sqrt{-g}F_{\alpha\beta}g^{\alpha\rho}g^{\beta\sigma}$, which completes the proof that the Lagrangians~(\ref{Lagr2}) and~(\ref{Lagr4}) are dynamically equivalent.

The purely affine formulation of electromagnetism alone, however, has one problem.
In the zero-field limit, where $F_{\mu\nu}=0$, the Lagrangian density~(\ref{Lagr2}) vanishes, thus making it impossible to apply Eqs.~(\ref{met1}) and~(\ref{met2}) to construct the metric tensor.
Therefore, there must exist (at least in the absence of matter) a {\it background} field that depends on the tensor $P_{\mu\nu}$ so that the metric tensor is well-defined.
The simplest possibility, supported by recent astronomical observations~\cite{acc1,acc2,const}, and thus a natural candidate, is the cosmological constant represented by the Eddington Lagrangian density~(\ref{Lagr1}).
Since it is impossible to find a purely affine Lagrangian that produces the Einstein equations in vacuum, as we mentioned in Sect.~\ref{secEdd}, and the cosmological constant guarantees that the metric structure in purely affine gravity is well-defined in the absence of matter (e.g., electromagnetic fields), we can regard spacetime with a cosmological constant as {\em purely affine vacuum}.
In the following sections we will combine the electromagnetic field and the cosmological constant in the purely affine formulation.

\section{Affine Einstein-Born-Infeld formulation}
\label{secWeak}

Since the metric structure in the purely affine Ferraris-Kijowski model of electromagnetism is not well-defined in the zero-field limit, we need to combine the electromagnetic field and the cosmological constant.
In the Lagrangian density~(\ref{Lagr2}) we used the determinant of the symmetrized Ricci tensor $P_{\mu\nu}$, multiplied by the simplest scalar containing the electromagnetic field tensor and $P_{\mu\nu}$.
As an alternative way to add the electromagnetic field into purely affine gravity, we can include the tensor $F_{\mu\nu}$ inside this determinant,\footnote{
We could add to the expression~(\ref{Lagr1}) the determinant of the electromagnetic field tensor, $\sqrt{\textrm{det}F_{\mu\nu}}$.
Such a term, however, is independent of the Ricci tensor and the metric tensor density given by Eq.~(\ref{met1}) would not couple to the electromagnetic field tensor $F_{\mu\nu}$.
Moreover, $\sqrt{\textrm{det}F_{\mu\nu}}=-\frac{1}{8}\epsilon^{\mu\nu\rho\sigma}F_{\mu\nu}F_{\rho\sigma}=-\frac{1}{4}(\epsilon^{\mu\nu\rho\sigma}F_{\rho\sigma}A_\nu)_{,\mu}$~\cite{Schr} is a total divergence and does not contribute to the field equations~\cite{LL2}.
}
constructing a purely affine version of the Einstein-Born-Infeld theory~\cite{BI,Motz,Vol}.
For $F_{\mu\nu}=0$, this construction reduces to the Eddington Lagrangian so the metric structure in the zero-field limit is well-defined.
Therefore it describes both the electromagnetic field and cosmological constant.
Let us consider the following Lagrangian density:
\begin{equation}
\textgoth{L}=\frac{1}{\kappa\Lambda}\sqrt{-\mbox{det}(P_{\mu\nu}+B_{\mu\nu})},
\label{LagrBIE}
\end{equation}
where 
\begin{equation}
B_{\mu\nu}=i\sqrt{\kappa\Lambda}F_{\mu\nu}
\label{def1}
\end{equation}
and $\Lambda>0$.
Let us also assume
\begin{equation}
|B_{\mu\nu}|\ll |P_{\mu\nu}|,
\label{appr1}
\end{equation}
where the bars denote the order of the largest (in magnitude) component of the corresponding tensor.\footnote{
The purpose of $i$ in the definition~(\ref{def1}) is to ensure that the energy density of the electromagnetic field appears in the Einstein equations with the correct sign.
The Lagrangian density~(\ref{LagrBIE}) is real.
}
Consequently, we can expand the Lagrangian density~(\ref{LagrBIE}) in small terms $B_{\mu\nu}$.
If $s_{\mu\nu}$ is a symmetric tensor and $a_{\mu\nu}$ is an antisymmetric tensor, the determinant of their sum is given by~\cite{Scho,Hlav}
\begin{equation}
\mbox{det}(s_{\mu\nu}+a_{\mu\nu})=\mbox{det}s_{\mu\nu}\biggl(1+\frac{1}{2}a_{\alpha\beta}a_{\rho\sigma}s^{\alpha\rho}s^{\beta\sigma}+\frac{\mbox{det}a_{\mu\nu}}{\mbox{det}s_{\mu\nu}}\biggr),
\label{det1}
\end{equation}
where the tensor $s^{\mu\nu}$ is reciprocal to $s_{\mu\nu}$.
If we associate $s_{\mu\nu}$ with $P_{\mu\nu}$ and $a_{\mu\nu}$ with $B_{\mu\nu}$, and neglect the last term in Eq.~(\ref{det1}), we obtain
\begin{equation}
\mbox{det}(P_{\mu\nu}+B_{\mu\nu})=\mbox{det}P_{\mu\nu}\biggl(1+\frac{1}{2}B_{\alpha\beta}B_{\rho\sigma}P^{\alpha\rho}P^{\beta\sigma}\biggr).
\label{det2}
\end{equation}
In the same approximation, the Lagrangian density~(\ref{LagrBIE}) becomes
\begin{equation}
\textgoth{L}=\frac{1}{\kappa\Lambda}\sqrt{-\mbox{det}P_{\mu\nu}}\biggl(1+\frac{1}{4}B_{\alpha\beta}B_{\rho\sigma}P^{\alpha\rho}P^{\beta\sigma}\biggr),
\label{det3}
\end{equation}
which is equal to the sum of the Lagrangian densities~(\ref{Lagr1}) and~(\ref{Lagr2}).

Equations~(\ref{met1}) and~(\ref{met2}) define the contravariant metric tensor,\footnote{The tensor density ${\sf g}^{\mu\nu}$ remains symmetric since only the symmetrized Ricci tensor enters the Lagrangian.} for which we find\footnote{We use the identity $\delta P^{\rho\sigma}=-\delta P_{\mu\nu}P^{\rho\mu}P^{\sigma\nu}$.}
\begin{eqnarray}
& & \sqrt{-g}g^{\mu\nu}=-\frac{1}{\Lambda}\biggl[P^{\mu\nu}\biggl(1+\frac{1}{4}B_{\alpha\beta}B_{\rho\sigma}P^{\alpha\rho}P^{\beta\sigma}\biggl) \nonumber \\
& & -P^{\alpha\beta}B_{\alpha\rho}B_{\beta\sigma}P^{\mu\rho}P^{\nu\sigma}\biggr]\sqrt{-\mbox{det}P_{\mu\nu}}.
\label{det4}
\end{eqnarray}
In the terms containing $B_{\mu\nu}$ and in the determinant\footnote{
In our approximation, $\sqrt{-\mbox{det}P_{\mu\nu}}=\Lambda^2\sqrt{-g}$,
since, as we show below, $P_{\mu\nu}=-\Lambda g_{\mu\nu}+\kappa T_{\mu\nu}$ which yields $\mbox{det}P_{\mu\nu}-\mbox{det}(\Lambda g_{\mu\nu})\propto T_{\mu\nu}g^{\mu\nu}=0$.
}
we can use the relation $P^{\mu\nu}=-\Lambda^{-1}g^{\mu\nu}$ (equivalent to Eq.~(\ref{cosm3})) valid for $B_{\mu\nu}=0$.
As a result, we obtain
\begin{equation}
g^{\mu\nu}=-\Lambda P^{\mu\nu}+\Lambda^{-2}\biggl(\frac{1}{4}g^{\mu\nu}B_{\rho\sigma}B^{\rho\sigma}-B^{\mu\rho}B^\nu_{\phantom{\nu}\rho}\biggr).
\label{det5}
\end{equation}
Introducing the energy-momentum tensor for the electromagnetic field:
\begin{equation}
T^{\mu\nu}=\frac{1}{4}g^{\mu\nu}F_{\rho\sigma}F^{\rho\sigma}-F^{\mu\rho}F^\nu_{\phantom{\nu}\rho},
\label{emt}
\end{equation}
turns Eq.~(\ref{det5}) into
\begin{equation}
P^{\mu\nu}=-\Lambda^{-1}g^{\mu\nu}-\kappa\Lambda^{-2}T^{\mu\nu},
\label{det6}
\end{equation}
which is equivalent, in the approximation~(\ref{appr1}), to the Einstein equations of general relativity with the cosmological constant in the presence of the electromagnetic field:
\begin{equation}
P_{\mu\nu}=-\Lambda g_{\mu\nu}+\kappa T_{\mu\nu}.
\label{det7}
\end{equation}
As in the case for the gravitational field only, $P_{\mu\nu}=R_{\mu\nu}(g)$.
The tensor~(\ref{emt}) is traceless, from which it follows that
\begin{equation}
R_{\mu\nu}(g)-\frac{1}{2}R(g)g_{\mu\nu}=\Lambda g_{\mu\nu}+\kappa T_{\mu\nu}.
\label{det8}
\end{equation}

Since the tensor $R_{\mu\nu}(g)$ satisfies the contracted Bianchi identities:
\begin{equation}
\biggl(R_{\mu\nu}(g)-\frac{1}{2}R(g)g_{\mu\nu}\biggr)^{;\nu}=0,
\label{Bia}
\end{equation}
and $g_{\mu\nu}^{\phantom{\mu\nu};\nu}=0$ because of Eq.~(\ref{Chr}), the tensor $T_{\mu\nu}$ that appears in Eq.~(\ref{det7}) is covariantly conserved, $T_{\mu\nu}^{\phantom{\mu\nu};\nu}=0$.
As in the metric formulation of gravitation, this conservation results from the invariance of the action integral under the coordinate transformations~\cite{Schr}.

To obtain the Maxwell equations in vacuum, we vary the action integral of the Lagrangian density~(\ref{LagrBIE}) with respect to the electromagnetic potential and use the principle of least action for an arbitrary variation $\delta A_\mu$:
\begin{equation}
\delta I=\frac{1}{c}\int d^4x\frac{\partial\textgoth{L}}{\partial F_{\mu\nu}}\delta(A_{\nu,\mu}-A_{\mu,\nu})=0,
\label{Max1a}
\end{equation}
from which it follows that $(\frac{\partial\textgoth{L}}{\partial F_{\mu\nu}})_{,\mu}=0$, or
\begin{equation}
\biggl(\sqrt{-\mbox{det}P_{\alpha\beta}}F_{\rho\sigma}P^{\mu\rho}P^{\nu\sigma}\biggr)_{,\mu}=0.
\label{Max2a}
\end{equation}
In our approximation, where $P^{\mu\nu}\propto g^{\mu\nu}$, Eq.~(\ref{Max2a}) takes the form of Eqs.~(\ref{Max2}) and~(\ref{Max3}) which are consistent with the Bianchi identities~(\ref{Bia}) applied to Eqs.~(\ref{emt}) and~(\ref{det8}).

Equations~(\ref{Max2}) and~(\ref{det8}) are the Einstein-Maxwell equations derived from the purely afine Lagrangian~(\ref{LagrBIE}) in the approximation~(\ref{appr1}) which can also be written as
\begin{equation}
\frac{\kappa F^2}{\Lambda}\ll 1,
\label{appr2}
\end{equation}
where $F=|F_{\mu\nu}|$.
The magnetic field of the Earth is on the order of $10^{-5}\,T$, which gives $\frac{\kappa F^2}{\Lambda}\sim10^6$.
The threshold at which the approximation~(\ref{appr2}) ceases to hold occurs at the level of the magnetic field on the order of $10^{-8}\,T$, i.e. in outer space of the Solar System.
Thus, the model of gravitation and electromagnetism presented in this section is not valid for electromagnetic fields observed in common life, e.g., that of a small bar magnet, which is on the order of 0.01 $T$.\footnote{For these fields, $|B_{\mu\nu}|\gg |P_{\mu\nu}|$, and the Lagrangian density is approximately proportional to $\sqrt{\textrm{det}F_{\mu\nu}}$ which cannot describe electromagnetism.}
This model is valid only for magnetic fields in interstellar space on the order of $10^{-10}\,T$, for which $\frac{\kappa F^2}{\Lambda}\sim10^{-5}$, thus is not physical.

\section{Eddington plus Ferraris-Kijowski Lagrangian}
\label{secEddFK}

We now combine the results of Sect.~\ref{secEdd} and Sect.~\ref{secFK} to construct a purely affine model of the gravitational field produced by the electromagnetic field and cosmological constant.
In general relativity, the corresponding metric-affine (or metric) Lagrangian density is the sum of the two densities~(\ref{cosm4}) and~(\ref{Lagr4}):
\begin{equation}
\textgoth{H}_{\textrm{\scriptsize{Max}}+\Lambda}=\textgoth{H}_{\textrm{\scriptsize{Max}}}+\textgoth{H}_{\Lambda}=-\frac{1}{4}\sqrt{-g}F_{\alpha\beta}F_{\rho\sigma}g^{\alpha\rho}g^{\beta\sigma}-\frac{\Lambda}{\kappa}\sqrt{-g},
\label{Lagr5}
\end{equation}
where $\Lambda>0$.
From Eq.~(\ref{Lagr5}) it follows that
\begin{equation}
P_{\mu\nu}=\kappa\biggl(\frac{1}{4}F_{\alpha\beta}F_{\rho\sigma}g^{\alpha\rho}g^{\beta\sigma}g_{\mu\nu}-F_{\mu\alpha}F_{\nu\beta}g^{\alpha\beta}\biggr)-\Lambda g_{\mu\nu},
\label{EinMax2}
\end{equation}
which yields $P=-4\Lambda$.
Therefore, Eq.~(\ref{Leg1}) reads
\begin{equation}
\textgoth{L}_{\textrm{\scriptsize{Max}}+\Lambda}=\textgoth{H}_{\textrm{\scriptsize{Max}}+\Lambda}+\frac{2\Lambda}{\kappa}.
\label{dev1}
\end{equation}
In the frame of reference in which Eqs.~(\ref{Gal}) and~(\ref{EM}) hold, we find
\begin{equation}
P_{\mu\nu}=\left( \begin{array}{cccc}
k-\Lambda & 0 & 0 & 0 \\
0 & \Lambda-k & 0 & 0 \\
0 & 0 & k+\Lambda & 0 \\
0 & 0 & 0 & k+\Lambda \end{array} \right),
\label{REM7}
\end{equation}
where we define
\begin{equation}
k=\frac{\kappa}{2}(E^2+B^2).
\label{ka}
\end{equation}
Consequently, we obtain
\begin{equation}
\mbox{det}P_{\mu\nu}=-(k^2-\Lambda^2)^2
\label{detP}
\end{equation}
and
\begin{eqnarray}
& & \sqrt{-\mbox{det}P_{\mu\nu}}F_{\alpha\beta}P^{\alpha\rho}P^{\beta\sigma}=\mbox{sign}(k^2-\Lambda^2) \nonumber \\
& & \times\left( \begin{array}{cccc}
0 & -E\frac{k+\Lambda}{k-\Lambda} & 0 & 0 \\
E\frac{k+\Lambda}{k-\Lambda} & 0 & 0 & 0 \\
0 & 0 & 0 & -B\frac{k-\Lambda}{k+\Lambda} \\
0 & 0 & B\frac{k-\Lambda}{k+\Lambda} & 0 \end{array} \right).
\label{no}
\end{eqnarray}

Let us construct the purely affine Lagrangian density for the electromagnetic field and gravity with cosmological constant as the simple sum of the two densities~(\ref{Lagr1}) and~(\ref{Lagr2}):
\begin{eqnarray}
& & \textgoth{L}_{\textrm{\scriptsize{FK}}+\textrm{\scriptsize{Edd}}}=\textgoth{L}_{\textrm{\scriptsize{FK}}}+\textgoth{L}_{\textrm{\scriptsize{Edd}}} \nonumber \\
& & =-\frac{1}{4}\sqrt{-\mbox{det}P_{\mu\nu}}F_{\alpha\beta}F_{\rho\sigma}P^{\alpha\rho}P^{\beta\sigma}+\frac{1}{\kappa\Lambda}\sqrt{-\mbox{det}P_{\mu\nu}}.
\label{Lagr6}
\end{eqnarray}
For this density, Eq.~(\ref{met1}) yields
\begin{eqnarray}
& & {\sf g}^{\mu\nu}=-\frac{1}{\Lambda}\sqrt{-\mbox{det}P_{\rho\sigma}}P^{\mu\nu}+\kappa\sqrt{-\mbox{det}P_{\rho\sigma}}P^{\beta\sigma} \nonumber \\
& & \times F_{\alpha\beta}F_{\rho\sigma}\biggl(\frac{1}{4}P^{\mu\nu}P^{\alpha\rho}-P^{\mu\alpha}P^{\nu\rho}\biggr).
\label{extra2}
\end{eqnarray}
The Lagrangian density~(\ref{Lagr6}) is identical with the (electromagnetic) weak-field approximation~(\ref{det3}) of the Lagrangian~(\ref{LagrBIE}).
Equations~(\ref{REM7}), (\ref{no}) and (\ref{Lagr6}) yield
\begin{equation}
\textgoth{L}_{\textrm{\scriptsize{FK}}+\textrm{\scriptsize{Edd}}}=\mbox{sign}(k^2-\Lambda^2)\times\biggl[\frac{1}{2}\biggl(E^2\frac{k+\Lambda}{k-\Lambda}-B^2\frac{k-\Lambda}{k+\Lambda}\biggr)+\frac{1}{\kappa\Lambda}(\Lambda^2-k^2)\biggr].
\label{Lagr7}
\end{equation}

We see that the expression~(\ref{Lagr7}) has a singular behavior in the limit $\Lambda\rightarrow0$ when $k\neq0$.
Therefore, the sum of the purely affine Ferraris-Kijowski and Eddington Lagrangians cannot describe, unlike the Einstein-Maxwell-$\Lambda$ theory, electromagnetic fields without the cosmological constant.
Moreover, from the relation
\begin{equation}
\textgoth{H}_{\textrm{\scriptsize{Max}}+\Lambda}=\frac{1}{2}(E^2-B^2)-\frac{\Lambda}{\kappa},
\label{dev2}
\end{equation}
it follows that if $k\neq0$ then
\begin{equation}
\textgoth{L}_{\textrm{\scriptsize{FK}}+\textrm{\scriptsize{Edd}}}-\textgoth{H}_{\textrm{\scriptsize{Max}}+\Lambda}-\frac{2\Lambda}{\kappa}\neq0.
\label{dev3}
\end{equation}
Comparing Eq.~(\ref{dev3}) with Eq.~(\ref{dev1}) indicates that the affine Lagrangian density $\textgoth{L}_{\textrm{\scriptsize{FK}}+\textrm{\scriptsize{Edd}}}$ is dynamically {\em inequivalent} to the metric-affine (or metric) Lagrangian density $\textgoth{H}_{\textrm{\scriptsize{Max}}+\Lambda}$ unless $k=0$ (this inequivalence was shown in Ref.~\cite{Proceed} by finding the purely affine equivalent of the Einstein-Maxwell-$\Lambda$ Lagrangian, see the next section).
In other words:
\begin{equation}
\textgoth{L}=\textgoth{L}_{\textrm{\scriptsize{FK}}+\textrm{\scriptsize{Edd}}}\Rightarrow\textgoth{H}\neq\textgoth{H}_{\textrm{\scriptsize{Max}}+\Lambda}
\end{equation}
and
\begin{equation}
\textgoth{H}=\textgoth{H}_{\textrm{\scriptsize{Max}}+\Lambda}\Rightarrow\textgoth{L}\neq\textgoth{L}_{\textrm{\scriptsize{FK}}+\textrm{\scriptsize{Edd}}}.
\end{equation}
The dynamics governed by the affine Lagrangian density $\textgoth{L}_{\textrm{\scriptsize{FK}}+\textrm{\scriptsize{Edd}}}$~(\ref{Lagr6}) deviates significantly from the dynamics governed by the experimentally and observationally verified metric-affine (or metric) Lagrangian density $\textgoth{H}_{\textrm{\scriptsize{Max}}+\Lambda}$ (\ref{Lagr5}) (or its purely affine equivalent, see Sect.~\ref{secPA}) unless $k\ll\Lambda$.
Therefore the Lagrangian density~(\ref{Lagr6}) is valid only for very weak electromagnetic fields, like the Lagrangian density~(\ref{LagrBIE}), and thus {\em unphysical}.

If the metric-affine (or metric) Lagrangian for the electromagnetic field and cosmological constant is the simple sum of the Einstein-Maxwell and $\Lambda$ Lagrangians, then the dynamically equivalent affine Lagrangian will be more complicated~\cite{KW}.
Similarly, if we assume that the affine Lagrangian for the electromagnetic field and cosmological constant is the simple sum of the Ferraris-Kijowski and Eddington Lagrangian, then the corresponding metric-affine (or metric) Lagrangian will be more complicated.\footnote{
The above results do not change if we add massive objects and restrict our analysis to the gravitational field outside them.}
This result is not surprising because the purely affine Lagrangian for the Einstein-Klein-Gordon theory already shows that what is a sum in the metric picture:
\begin{equation}
\textgoth{H}=\sqrt{-g}\Bigl(\frac{1}{2}\phi_{,\mu}\phi_{,\nu}g^{\mu\nu}-\frac{1}{2}m^2\phi^2\Bigr),
\label{KG1}
\end{equation}
 is not a sum in the affine picture~\cite{Kij,KW}:
\begin{equation}
\textgoth{L}=\frac{2}{m^2\phi^2}\sqrt{\Bigl|\mbox{det}\Bigl(\frac{P_{\mu\nu}}{\kappa}-\phi_{,\mu}\phi_{,\nu}\Bigr)\Bigr|}.
\label{KG2}
\end{equation}
The dynamical inequivalence between the two (purely affine and metric-affine/metric) simple Lagrangians indicates that the Ferraris-Kijowski model of electromagnetism~\cite{FK1}, modified additively by the term that ensures a well-defined zero-field limit, physically deviates from the Einstein-Maxwell-$\Lambda$ equations (electromagnetic fields and dark energy will interact) and can be tested~\cite{Col}.
These deviations may be significant for systems with strong electromagnetic fields, such as neutron stars.

To illustrate this point further, we can use an analogy with classical mechanics.
Let us consider a simple Lagrangian $L_\alpha=\frac{\alpha}{2}\dot{q}^2$, where $\alpha$ does not depend on $q$.
The corresponding Hamiltonian $H_\alpha=\frac{p^2}{2\alpha}$ is simple.
Let us now consider another simple Lagrangian $L_\beta=\frac{\beta}{3}\dot{q}^3$, where $\beta$ does not depend on $q$.
The corresponding Hamiltonian $H_\beta=\frac{2}{3}\sqrt{\frac{p^3}{\beta}}$ is simple too.
However, if we take the sum of the two above Lagrangians, $L=L_\alpha+L_\beta$, the corresponding Hamiltonian is
\begin{equation}
H=\frac{1}{12\beta^2}((\alpha^2+4\beta p)^{3/2}-\alpha^3)-\frac{\alpha p}{2\beta},
\label{ex1}
\end{equation}
which differs from the simple expression $H=H_\alpha+H_\beta$.
This Hamiltonian reduces to $H_\alpha$ if $\beta=0$ and to $H_\beta$ if $\alpha=0$.
Conversely, one can show that the Lagrangian corresponding to the sum of the two simple Hamiltonians, $H=H_\alpha+H_\beta$, differs from $L=L_\alpha+L_\beta$.
If we regard $\alpha$ as a quantity representing the cosmological constant and $\beta$ as a quantity representing the electromagnetic field, it is clear why a simple (additive with respect to $\alpha$ and $\beta$) Lagrangian density in one picture (affine or metric-affine/metric) is not simple in the other ($\alpha$ and $\beta$ {\em interact}).

\section{Purely affine equivalent of Einstein-Maxwell-$\Lambda$ Lagrangian}
\label{secPA}

The purely affine Lagrangian that is dynamically equivalent to the purely metric Einstein-Maxwell Lagrangian with the cosmological constant~(\ref{Lagr5}) was found in Ref.~\cite{Proceed}.
Here we briefly show the intermediate steps leading to this result.
Equation~(\ref{EM}) gives
\begin{equation}
\mbox{det}F_{\mu\nu}=E^2 B^2.
\label{1}
\end{equation}
Equations~(\ref{Gal}), (\ref{EM}), (\ref{Lagr5}) and~(\ref{dev1}) give
\begin{equation}
\textgoth{L}_{\textrm{\scriptsize{Max}}+\Lambda}=\frac{1}{2}(E^2 - B^2)+\frac{\Lambda}{\kappa}.
\label{2}
\end{equation}
Combining Eqs.~(\ref{1}) and~(\ref{2}) gives
\begin{equation}
4\kappa\Lambda(\textgoth{L}^2_{\textrm{\scriptsize{Max}}+\Lambda}+\mbox{det}F_{\mu\nu})=\kappa\Lambda(E^2 + B^2)^2 + 4\Lambda^2(E^2 - B^2) + \frac{4\Lambda^3}{\kappa}.
\label{3}
\end{equation}
From (\ref{detP}) we also have
\begin{equation}
\sqrt{-\mbox{det}P_{\mu\nu}}\mbox{sign}(k^2 - \Lambda^2)=\frac{\kappa^2}{4}(E^2 + B^2)^2 - \Lambda^2,
\label{4}
\end{equation}
which, with Eq.~(\ref{2}), gives
\begin{eqnarray}
& & -4\sqrt{-\mbox{det}P_{\mu\nu}}\mbox{sign}(k^2 - \Lambda^2)\textgoth{L}_{\textrm{\scriptsize{Max}}+\Lambda}=-\frac{\kappa^2}{2}(E^2 - B^2)(E^2 + B^2)^2 \nonumber \\
& & - \kappa\Lambda(E^2 + B^2)^2 + 2\Lambda^2(E^2 - B^2) + \frac{4\Lambda^3}{\kappa}.
\label{5}
\end{eqnarray}
Finally, we find
\begin{eqnarray}
& & \mbox{det}(P_{\mu\nu})F_{\alpha\beta}F_{\rho\sigma}P^{\alpha\rho}P^{\beta\sigma}=\frac{\kappa^2}{2}(E^2 - B^2)(E^2 + B^2)^2 \nonumber \\
& & + 2\Lambda^2(E^2 - B^2) + 2\kappa\Lambda(E^2 + B^2)^2.
\label{6}
\end{eqnarray}

The left-hand side of Eq.~(\ref{3}) minus the sum of the left-hand sides of Eqs.~(\ref{5}) and~(\ref{6}) equals identically zero for arbitrary values of $E$, $B$ and $\Lambda$:
\begin{eqnarray}
& & 4\kappa\Lambda(\textgoth{L}^2_{\textrm{\scriptsize{Max}}+\Lambda}+\mbox{det}F_{\mu\nu})+4\sqrt{-\mbox{det}P_{\mu\nu}}\mbox{sign}(k^2 - \Lambda^2)\textgoth{L}_{\textrm{\scriptsize{Max}}+\Lambda} \nonumber \\
& & -\mbox{det}(P_{\mu\nu})F_{\alpha\beta}F_{\rho\sigma}P^{\alpha\rho}P^{\beta\sigma}=0,
\label{7}
\end{eqnarray}
from which we obtain the solution for $\textgoth{L}_{\textrm{\scriptsize{Max}}+\Lambda}$~\cite{Proceed}:
\begin{eqnarray}
& & \textgoth{L}_{\textrm{\scriptsize{Max}}+\Lambda}=\frac{1}{2\kappa\Lambda}\Bigl[-\mbox{sign}(k^2 - \Lambda^2)\sqrt{-\mbox{det}P_{\mu\nu}}+\Bigl(-\mbox{det}P_{\mu\nu} \nonumber \\
& & +\kappa\Lambda\,\mbox{det}(P_{\mu\nu})F_{\alpha\beta}F_{\rho\sigma}P^{\alpha\rho}P^{\beta\sigma}-4\kappa^2\Lambda^2\mbox{det}F_{\mu\nu}\Bigl)^{1/2}\Bigr].
\label{8}
\end{eqnarray}
In the limit of zero electromagnetic field, $F_{\mu\nu} \rightarrow0$, we have $\mbox{sign}(k^2 - \Lambda^2)=-1$ and
\begin{equation}
\textgoth{L}_{\textrm{\scriptsize{Max}}+\Lambda}\rightarrow \frac{1}{\kappa\Lambda}\sqrt{-\mbox{det}P_{\mu\nu}},
\label{9}
\end{equation}
reproducing the Eddington Lagrangian~(\ref{Lagr1}).
In the limit of zero cosmological constant, $\Lambda\rightarrow0$,\footnote{
The limit $\Lambda\rightarrow0$ is meaningful for the Lagrangian density~(\ref{8}), but not for~(\ref{Lagr6}).
}
we have $\mbox{sign}(k^2 - \Lambda^2)=1$ and
\begin{equation}
\textgoth{L}_{\textrm{\scriptsize{Max}}+\Lambda}\rightarrow -\frac{1}{4}\sqrt{-\mbox{det}P_{\mu\nu}}F_{\alpha\beta}F_{\rho\sigma}P^{\alpha\rho}P^{\beta\sigma},
\label{10}
\end{equation}
reproducing the Ferraris-Kijowski Lagrangian~(\ref{Lagr2}).
Since the Lagrangian densities (\ref{Lagr6}) and (\ref{8}) are quite different, it is clear why the sum of the purely affine Ferraris-Kijowski and Eddington Lagrangians does not describe, unlike the Einstein-Maxwell-$\Lambda$ theory, physical systems.

We can remove the term with $\mbox{det}F_{\mu\nu}$ from the affine Lagrangian density~(\ref{8}), simplifying this Lagrangian without changing its $F\rightarrow0$ and $\Lambda\rightarrow0$ limits.
The corresponding theory would deviate from the Einstein-Maxwell-$\Lambda$CDM theory (the electromagnetic field and the cosmological constant would interact) only for electromagnetic fields that contribute to the spacetime curvature on the same order as the cosmological constant, i.e. on the order of $10^{-8}\,T$.
The magnetic field in outer space of the Solar System, where we observe the Pioneer anomaly, is of this order~\cite{Pion}.

\section{Summary}
\label{secSum}

The purely affine formulation of gravity, in which the affine connection is a dynamical variable and the Lagrangian density depends on the symmetric part of the Ricci tensor of the connection, is physically equivalent to the metric-affine and purely metric formulation.
Therefore purely affine gravity is simply Einstein's general relativity formulated with different variables, analogously to Lagrangian mechanics being Hamiltonian mechanics formulated with generalized velocities instead of canonical momenta.
The equivalence of a purely affine gravity with general relativity, which is a metric theory, implies that the former is consistent with experimental tests of the weak equivalence principle~\cite{Wi}.

For each purely affine Lagrangian density we can construct a metric-affine matter Lagrangian density (which we call a Hamiltonian density) that is dynamically equivalent~\cite{FK3a,FK3b}.
If a purely affine Lagrangian depends on the connection only through the symmetric part of the Ricci tensor, the corresponding metric-affine and metric matter Lagrangians coincide.
The $\Lambda$CDM metric-affine (or metric) Lagrangian for the cosmological constant is dynamically equivalent to the Eddington affine Lagrangian.
The Einstein-Maxwell metric-affine (or metric) Lagrangian for the electromagnetic field is dynamically equivalent to the Ferraris-Kijowski affine Lagrangian, except the zero-field limit, where the metric tensor is not well-defined.
This feature indicates that, for the Ferraris-Kijowski model to be physical, there must exist a background field that depends on the Ricci tensor.
The simplest possibility, supported by recent astronomical observations, and thus a natural candidate, is the cosmological constant.
Therefore the purely affine formulation gives the cosmological constant the meaning of the entity assuring that all Lagrangians are well-defined in the zero-field limit.
Another advantage of purely affine gravity is a causal structure free of defects contained in the metric formulation~\cite{KW}.

In this paper we studied the combined electromagnetic field and cosmological constant in the purely affine formulation.
We found that the sum of the metric Maxwell and Einstein Lagrangian is dynamically inequivalent to the sum of the affine Ferraris-Kijowski and Eddington Lagrangian, which is also a direct consequence of a significant difference between the affine Lagrangians~(\ref{Lagr6}) and~(\ref{8}).
The results of Sect.~\ref{secEddFK} show that the sum of the purely affine Ferraris-Kijowski and Eddington Lagrangian~(\ref{Lagr6}) does not correspond to physical systems.
Therefore, one cannot apply the same rules that hold for metric theories to construct physically meaningful purely affine Lagrangians.
We also found that the affine Einstein-Born-Infeld formulation is not realistic for almost all electromagnetic fields existing in Nature.
The case of the affine equivalent of the metric Einstein-Maxwell-$\Lambda$ Lagrangian~(\ref{8}) shows that one has to be careful in choosing purely affine Lagrangians for physical theories.
The transformation of a purely metric Lagrangian into its dynamically equivalent affine Lagrangian turns out to be nonlinear.
 
A quite complicated form of the affine equivalent of the metric Einstein-Maxwell-$\Lambda$ Lagrangian does not indicate that the purely affine formulation of gravity is less physical than the metric one.
One can, in principle, construct an infinite number of purely affine Lagrangian for gravity, electromagnetism and cosmological constant that reproduces the Eddington Lagrangian in the limit of zero electromagnetic field and the Ferraris-Kijowski Lagrangian in the limit of zero cosmological constant.
If one of such affine Lagrangians is simple, we could consider it as a basis of a theory of gravity and electromagnetism in the presence of cosmological constant.
For example, removing the term with $\mbox{det}F_{\mu\nu}$ from the affine Lagrangian density~(\ref{8}) simplifies this Lagrangian without changing its $F\rightarrow0$ and $\Lambda\rightarrow0$ limits.
The corresponding theory would deviate from the standard Einstein-Maxwell-$\Lambda$CDM theory only for electromagnetic fields that contribute to the spacetime curvature on the same order as the cosmological constant.
The magnetic field in outer space of the Solar System is of this order, which suggests a possible explanation of the Pioneer anomaly by the purely affine formulation of gravity.

\begin{acknowledgments}
The author wishes to thank Prof. Marco Ferraris for providing Ref.~\cite{Proceed} and Prof. Jerzy Kijowski for recommending Ref.~\cite{KW}.
Both papers were very helpful in revising the present manuscript.
The author also thanks the anonymous referees for their helpful comments and suggestions.
\end{acknowledgments}



\end{document}